\documentclass[submission,copyright,creativecommons]{eptcs}
\providecommand{\event}{FROM 2024} 

\usepackage{iftex}

\ifpdf
  \usepackage{underscore}         
  \usepackage[T1]{fontenc}        
\else
  \usepackage{breakurl}           
\fi

\usepackage[utf8]{inputenc}
\usepackage[english]{babel}
\usepackage{url}
\usepackage{amssymb}
\usepackage{amsmath}
\usepackage{graphicx}
\usepackage{amsthm}

\newcommand{\Nat}{\ensuremath{\mathbb{N}}}

\newcommand{\Rel}{\ensuremath{\mathbb{R}}}

\newcommand{\bag}[1]{\ensuremath{Bag[#1]}}

\usepackage{listings}
\usepackage{xcolor}

\definecolor{delimiterColor}{HTML}{B65E47}
\definecolor{numberColor}{HTML}{FF0000}
\definecolor{commentColor}{HTML}{008000}
\definecolor{keyColor}{HTML}{002BFF}

\lstdefinelanguage{maude}
{
	breaklines=true,
	extendedchars=true,
	tabsize=2,
	columns=fullflexible,
	showtabs=false,
	showstringspaces=false,
	showspaces=false,
	showstringspaces=false,
	identifierstyle={\ttfamily},
	keywordstyle={\color{keyColor}},
	ndkeywordstyle={\color{keyColor}},
	stringstyle={\color{delimiterColor}},
	commentstyle={\color{commentColor}},
	ndkeywords={},
	keywords={pr, protecting, sort, sorts, op, ops, var, vars,eq, cq, ceq, crl, rl, mb, cmb, endfm, fmod, is, mod, endm, =, ==, =/=, ctor, ditto, Object, owise, Oid, prec, assoc, id, if, class, homod, endhom, eof, var, vars, eq, op, ops, pr, inc, protecting, including, ceq, is, tomod, endtom, sort, subsort, subsorts, to, endom, fmod, endfm, mod, endm, endtm, comm, gather, fth, endfth, format, metadata, memo},
	morecomment={[l]{***}},
	morecomment={[l]{---}},
}
\newtheorem{definition}{Definition}[section]

\newtheorem{property}{Property}[section]

\begin{document}

\title{Efficient Performance Analysis of Modular\\ Rewritable Petri Nets}

\author{Lorenzo Capra
\institute{
	{Dipartimento di Informatica}\\ {Universit{\`a} degli Studi di Milano}, Italy 
}
\and Marco Gribaudo
\institute{
{Dipartimento di Elettronica, Informatica e Bioingeneria}\\
{Politecnico di Milano}, Italy
}
}

\def\titlerunning{Performance analysis of RwPT}
\def\authorrunning{Capra, Gribaudo}

\maketitle

\begin{abstract}
Petri Nets (PN) are extensively used as a robust formalism to model concurrent and distributed systems; however, they encounter difficulties in accurately modeling adaptive systems. To address this issue, we defined rewritable PT nets (RwPT) using \texttt{Maude}, a declarative language that ensures consistent rewriting logic semantics. Recently, we proposed a modular approach that employs algebraic operators to build extensive RwPT models. This methodology uses composite node labeling to maintain hierarchical organization through net rewrites and has been shown to be effective.
Once stochastic parameters are integrated into the formalism, we introduce an automated procedure to derive a \emph{lumped} CTMC from the quotient graph generated by a modular RwPT model.
To demonstrate the effectiveness of our method, we present a fault-tolerant manufacturing system as a case study.


\end{abstract}


\section{Introduction}
\label{sec:intro}
Despite their potential, traditional formalisms such as Petri Nets, Automata, and Process Algebra do not easily allow designers to define dynamic changes in systems or assess their performance impact. As a result, many extensions to these classical models have been proposed, such as the $\pi$-calculus and the Nets-within-Nets paradigm, although they often lack adequate analysis techniques.

Rewritable PT nets (RwPT) were introduced in \cite{lcICDCIT22} as a versatile formalism for the modeling and analysis of adaptive distributed systems. The RwPT procedures were defined using the declarative language \texttt{Maude}, which leverages Rewriting Logic to offer both operational and mathematical semantics, thereby enabling a scalable framework for self-adapting PT nets. Unlike comparable methods (\cite{Barbosa11,RPN-Maude2016}), which convert a simpler type of PNs into \texttt{Maude}, the RwPT formalism simplifies data abstraction, is concise and effective, and circumvents the limitations imposed by the pushout mechanism common in Graph Transformation Systems. RwPT extends GTS. It is vital to consider graph isomorphism (GI) in recognizing equivalent states within the model dynamics. This consideration is particularly advantageous for scaling up the model's size or degree of parallelism, especially when integrating a Stochastic Process into the model's state space. Recent research has demonstrated that GI has a quasi-polynomial complexity \cite{GIBabai}. Graph canonization (GC), which is at least as complex as GI, involves determining a canonical form for any graph such that for any two graphs $G$ and $G'$, $G \simeq G' \Leftrightarrow can(G) = can(G')$. We have developed a general canonization method \cite{Capra:RP22} for use with RwPT, integrated into \texttt{Maude}. This method is effective for irregular models, but it is less efficient for more realistic models that contain numerous similar components organized in a nested structure.

In \cite{CAPRA-TCS2024}, we introduced a method for developing extensive RwPT models using algebraic operators. Our approach is simple: Exploiting the modular features of the models during analysis. Using composite node labeling, we identify symmetries and maintain hierarchical organization through net rewrites. A case study (used as benchmark) demonstrates the success of our method, showing performance benefits over somewhat related approaches. In this paper, we present an automated technique to derive a Lumped Continuous-Time Markov Chain (CTMC) from the quotient graph generated by an RwPT model after embedding stochastic parameters into the framework.

Background information is provided in Section \ref{sec:backgr}, and our example is described in Section \ref{sec:exe}. The modular RwPT formalism, now with stochastic parameters, is explained in Section \ref{sec:rewPT}. In Section \ref{sec:CTMC}, we detail the method for deriving a lumped CTMC from an RwPT model and present experimental evidence of its effectiveness. We conclude by discussing ongoing work.

\section{Background: (Stochastic) PT Nets and \texttt{Maude}}
\label{sec:backgr}

This section provides a concise overview of the (stochastic) PT formalism and emphasizes the key aspects of the 	\texttt{Maude} framework. For exhaustive information, we direct readers to the reference papers.

A \textit{multiset} (\textit{bag}) $b$ in a nonempty set $D$ is a map $b: D \rightarrow \Nat$, where $b(d)$ is the \emph{multiplicity} of $d$ in $b$. A multiset is empty if all of its multiplicities are zero. We denote by $Bag[D]$ the set of multisets in $D$.
Standard relational and arithmetic operations can be applied to multisets on a component-by-component basis.
In particular, let $b, b' \in Bag[D]$:

$b + b' \in Bag[D]$ is $b + b'(d) = b(d)+b'(d)$, $\forall d\in D$.

$b < b' = true $ $\Leftrightarrow$ $\forall d\in D \ b(d) < b'(d)$.

$b-b'\in Bag[D]$ is defined if $b' \leq b$: $b-b'(d) = b(d)-b'(d)$, $\forall d\in D$.
%

\medskip
A stochastic PT (or SPN) \emph{net}~\cite{ReisigPN,GSPN1993}
is a 6-tuple $(P,T,I,O,H,\lambda)$, where:
$P$, $T$ are finite, non-empty, disjoint sets holding the net's nodes (places and transitions, respectively);
$I,O,H:$ $T \ \rightarrow \bag{P}$ represent the transitions' \emph{input}, \emph{output}, and \emph{inhibitor} incidence matrices, respectively;
$\lambda : T \ \rightarrow \Rel^+$ assigns each transition a negative exponential firing rate. 
A PT net \emph{marking} is a multiset $m \in Bag[P]$.

The PT net dynamics
is defined by the \emph{firing rule}:
$t\in T$ is \emph{enabled} in marking $m$ if and only if: 
$$I(t) \leq m \wedge \forall p \in P : \ H(t)(p) > 0 \Rightarrow  m(p) < H(t)(p)$$ 

\noindent If $t$ is enabled $m$ it may fire, leading to marking 
$$m^\prime = m - I(t) + O(t)$$
The notation $m [ t \rangle m'$ means that $t$ is enabled in $m$ and its firing leads to $m'$.

A PT-\emph{system} is a pair $(N,m_0)$, where $N$ is a PT net and $m_0$ is a marking of $N$.
The interleaving semantics of $(N,m_0)$ is specified by the \emph{reachability graph} (RG): the RG is an edge-labelled, directed graph $(V, E)$  whose nodes are markings. It is defined inductively: $m_0 \in V$; if $m \in V$ and
$m [ t\rangle m'$
then $m' \in V$, $m \xrightarrow{t} m' \in E$.

The timed semantics of a stochastic PT system is a CTMC isomorphic to the RG.
For any two $m_i, m_j \in V$, the transition rate from $m_i$ to $m_j$ is $r_{i,j} := \sum_{t : m_i [ t \rangle m_j} \lambda(t)$. The CTMC infinitesimal generator is a $|V| \times |V|$ matrix $Q$ such that $Q[i,j] = r_{i,j}$ if $i \neq j$,
$Q[i,i] = 1 - \sum_{j, j \neq i} r_{i,j}$.

In Generalized Stochastic Petri Nets (GSPN) \cite{GSPN1993} transitions can be assigned a priority (the firing rule is extended accordingly): transitions with a priority greater than zero occur instantly, and the associated stochastic parameters (denoted by the function $\lambda$) are used to resolve potential conflicts probabilistically. Consequently, their timed semantics leads to a Continuous-Time Markov Chain (CTMC) that is isomorphic to the "reduced" RG, obtained by eliminating nonobservable markings. This paper focuses on Stochastic Petri Nets (SPN) even though our specification encompasses GSPN.

\paragraph{The \texttt{Maude} system}
\texttt{Maude} \cite{maude07} is a highly expressive, purely declarative language characterized by a rewriting logic semantics \cite{rewlog03}. Statements consist of (conditional) \emph{equations} and \emph{rules}. Each side of a rule or equation represents terms of a specific \emph{kind} that might include variables. The semantics of rules and equations involve straightforward rewriting, where instances of the left-hand side are substituted by corresponding instances of the right-hand side. The expressivity of \texttt{Maude} is realized through the use of matching modulo operator equational attributes, sub-typing, partiality, generic types, and reflection. A \texttt{Maude} \emph{functional} module comprises only \emph{equations} and functions as a functional program defining one or more operations through equations, utilized as simplification rules. A functional module details an \emph{equational theory} within membership equational logic \cite{membeqlog00}. Formally, such a theory is a tuple $(\Sigma,E \cup A)$, with $\Sigma$ representing the signature, which includes the declaration of all sorts, subsorts, kinds\footnote{Kinds are implicit equivalence classes defined by connected components of sorts (as per subsort partial order). Terms in a kind without a specific sort are \emph{error} terms.}, and operators; $E$ being the set of equations and membership axioms; and $A$ as the set of operator equational attributes (e.g., \texttt{assoc}). The model of $(\Sigma,E \cup A)$ is the \emph{initial algebra} $T_{\Sigma /E \cup A}$, which mathematically corresponds to the quotient of the ground-term algebra $T_{\Sigma}$. Provided that $E$ and $A$ satisfy nonrestrictive conditions, the final (or \emph{canonical}) values of ground terms form an algebra isomorphic to the initial algebra, ensuring that the mathematical and rewriting semantics are identical.

A \texttt{Maude} \emph{system module} includes \emph{rewrite rules} and, potentially, equations. These rules illustrate local transitions in a concurrent system. In formal language, a system module outlines a generalized \emph{rewrite theory} \cite{rewlog03}, symbolized as a four-tuple $\mathcal{R}= (\Sigma,E \cup A,\phi,R)$, where $(\Sigma,E \cup A)$ constitutes a membership equational theory; $\phi$ identifies the frozen arguments for each operator in $\Sigma$; and $R$ contains a set of rewrite rules \footnote{Rewrite rules do not apply to frozen arguments.}. This rewrite theory models a concurrent system. $(\Sigma,E \cup A)$ establishes the algebraic structure of the states, while $R$ and $\phi$ define the concurrent transitions of the system. The initial model of $\mathcal{R}$ assigns to each kind $k$ a labeled transition system (TS) where the states are the elements of $T_{\Sigma /E \cup A,k}$, and transitions occur as $ [t] \overset{[\alpha] }{\rightarrow} [t']$, with $[\alpha]$ representing \emph{equivalent} rewrites. The property of \emph{coherence} guarantees that a strategy that reduces terms to their canonical forms before applying the rules is sound and complete. A \texttt{Maude} system module is also an executable specification of distributed systems. Given finite reachability, it enables the verification of invariant properties and the discovery of counterexamples. Moreover, it supports the verification of LTL formulas. When the TS generated by a ground term becomes excessively large or infinite, bounded searches or abstractions might be employed.

\section{Gracefully Degrading Production System}
\label{sec:exe}

\begin{figure}[htbp]
   \begin{center}    
    \begin{tabular}{c}
        \fbox{\includegraphics[scale=0.6]{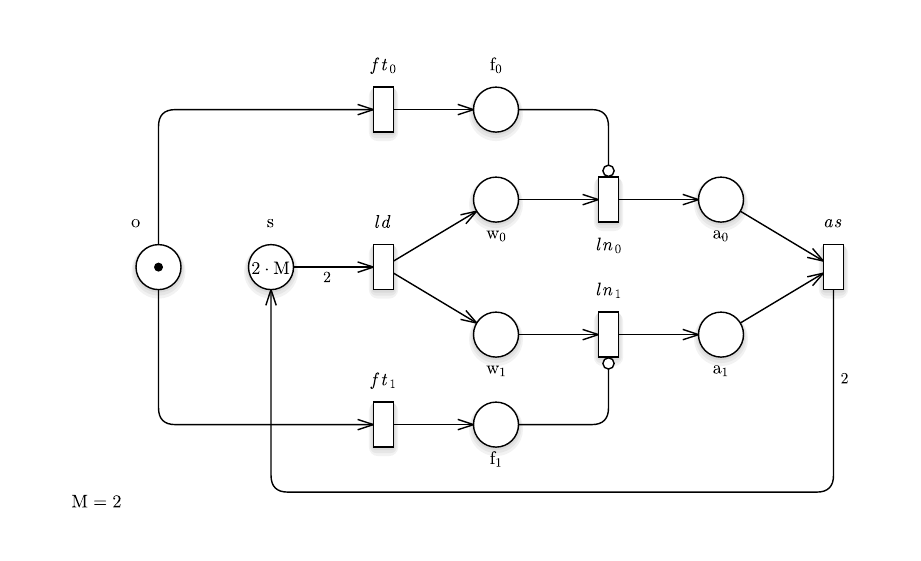}}\\
        $\Downarrow$\\

  \fbox{\includegraphics[scale=0.6]{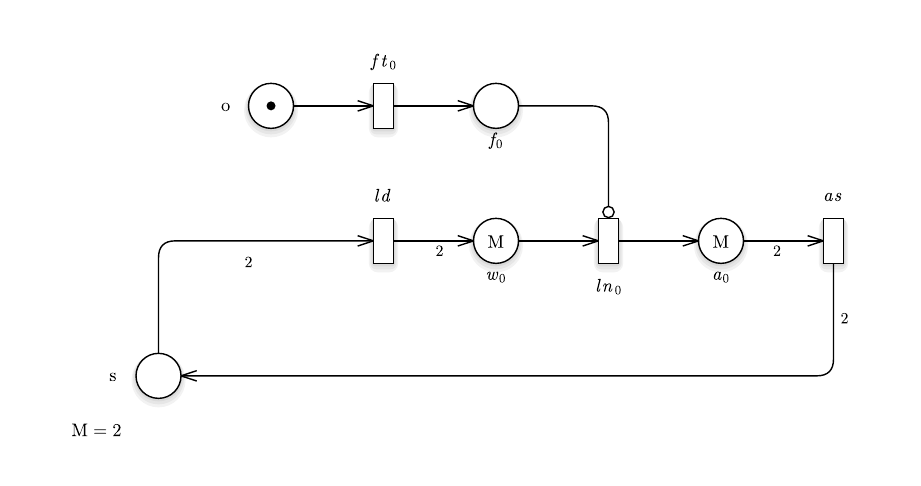}}
    \end{tabular}
    \end{center}
    \caption{Production Line (PL) and adaptation following a fault.}
    \label{fig:FMS}
\end{figure}

The illustrative example in this paper depicts a distributed production system that degrades gracefully, whose base configuration is shown by the two PT systems in Figure~\ref{fig:FMS}. The upper net represents a Production Line (denoted PL) which is divided into $K$ branches (robots) that handle raw materials (a multiple of $K$). These branches ($\{w_i, ln_i, a_i\}$, $i:0\ldots K-1$) are fully interchangeable. An assembly component (transition $as$) converts the processed materials $K$ into an artifact. A loader ($ld$) collects $K$ items from a storage facility (place $s$) on the $K$ lines of the PL. In this study, $K = 2$. The initial count of pieces (tokens) in $s$ is $K\cdot M$, where $M \in \Nat^+$ is another parameter of the model. For each artifact produced, fresh items $K$ are introduced. A branch might fail (transitions $ft_i$). When that occurs, the PL restructures to continue functioning, but with reduced capacity. Simple static analysis can show that the PL system reaches a \emph{deadlock} after a failure.
 
The net at the bottom of Figure~\ref{fig:FMS} shows the transformation of the PL after a fault happens (considering scenario $K = 2$). This process involves moving items from the faulted branch to the remaining branch(es) to maintain the production cycle. Traditional PN frameworks (including High-Level PN variants) are unable to model this operation. Items left on the faulty line (represented as place $w_1$ here) are transferred to the remaining functional line ($w_0$): The marking of the PT net at the bottom demonstrates the state after adaptation. We assume that a PL that fails twice is beyond repair.

We will examine a more complex scenario in which $N$ PL replicas function simultaneously and degrade in a regulated fashion. Figure \ref{fig:symPL-degarde2} demonstrates one potential evolution of a system starting with two PLs: This scenario can be extended to a system that incorporates $N$ PLs, each operating $K$ parallel robots, that handles $K \cdot M$ raw items, denoted by the term \verb|NPLsys(N, K, M)|. The graceful degradation of the system proceeds in two phases:

\begin{figure}[htbp]
   \begin{center}    
    \begin{tabular}{rl}
        \fbox{\includegraphics[scale=0.43]{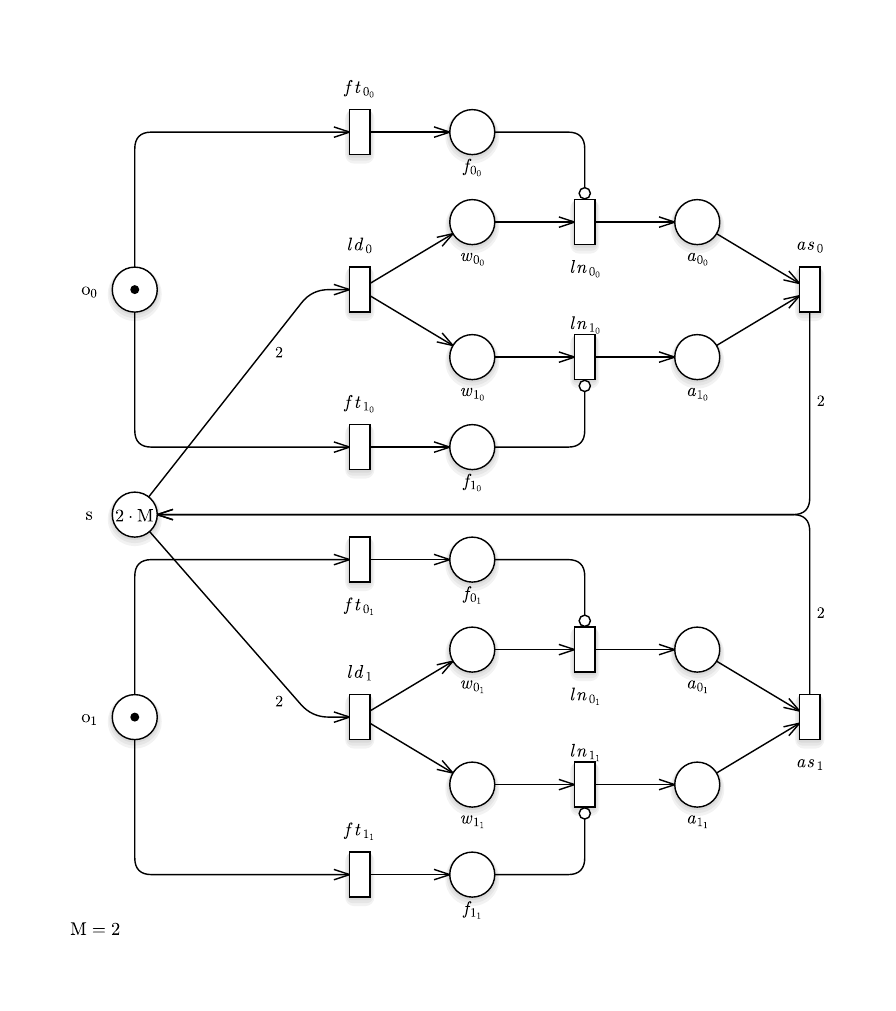}}&
        $\Rightarrow$
        \fbox{\includegraphics[scale=0.43]{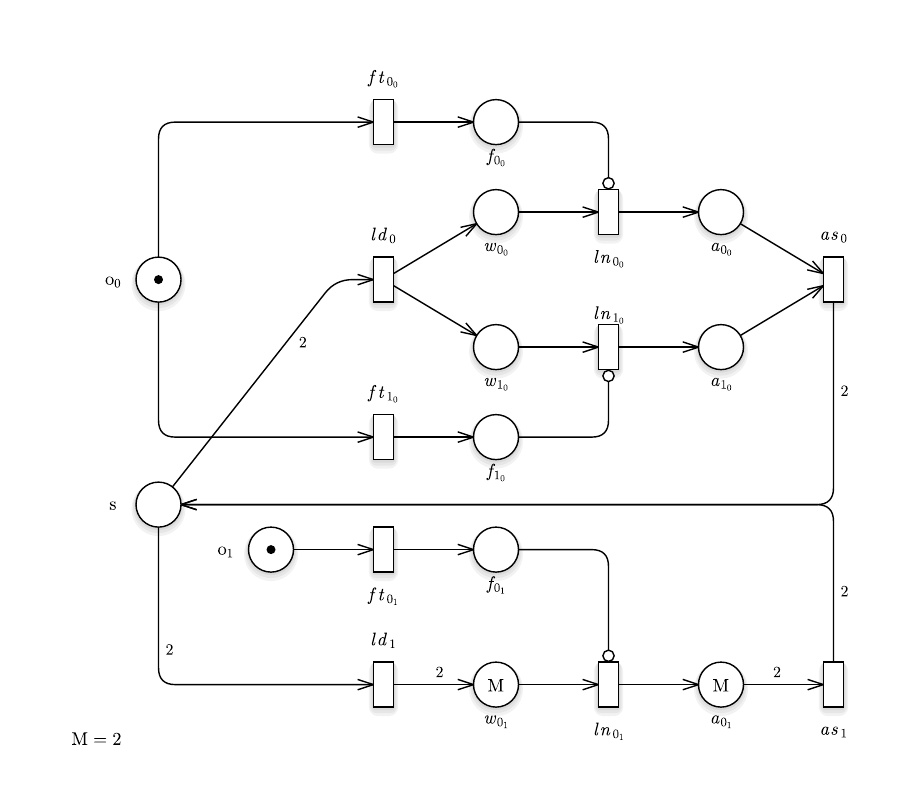}}
        $\Rightarrow$\\
        $~$&$~$\\
        $\Rightarrow$
        \fbox{\includegraphics[scale=0.43]{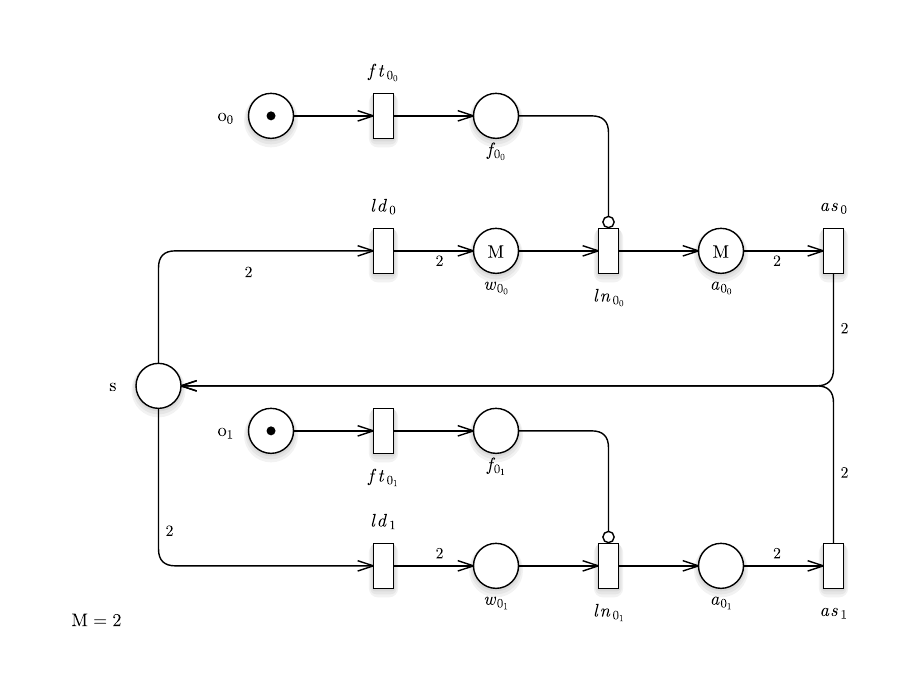}}&
        $\Rightarrow$
        \fbox{\includegraphics[scale=0.43]{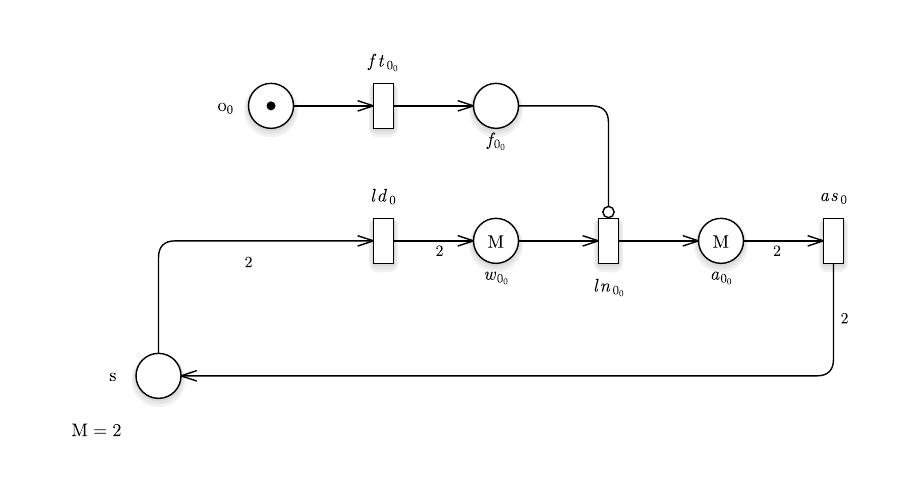}}
    \end{tabular}
    \end{center}
    \caption{One of the possible paths of the Gracefully Degrading Production System.}
    \label{fig:symPL-degarde2}
\end{figure}

\begin{itemize}
    \item[s1] When a fault impacts a robot (line) of a PL, the PL autonomously adjusts to continue functioning in a diminished capacity (for simplicity, we here consider a two-lines scenario)
    \item[s2] When a second fault occurs in a degraded PL, the PL is disconnected from the entire system (see the final step in Figure \ref{fig:symPL-degarde2}): The leftover items are then relocated to the warehouse.
\end{itemize}

\section{Modular Rewritable Stochastic PN}
\label{sec:rewPT}
This section introduces the concept of rewritable stochastic PT nets (RwSPT). These expand on the \emph{modular} rewritable PT nets described in \cite{CAPRA-TCS2024} by incorporating transitions with priorities and stochastic parameters. An RwSPT serves as an algebraic model of a Generalized Stochastic PN \cite{GSPN1993}, combining rewrite rules with the PT firing mechanism. In this study, we concentrate on stochastic PN consisting of zero-priority transitions accompanied by an exponential firing rate.

The definition of RwSPT includes a hierarchy of \textbf{Maude} modules (e.g., \texttt{BAG}, \texttt{PT-NET}, \texttt{PT-SYSTEM}) most of which described in  \cite{CAPRA-TCS2024}.
The \texttt{Maude} sources can be found in \url{https://github.com/lgcapra/rewpt/tree/main/modSPT}.

The RwSPT definition uses structured annotations to underline the model's symmetry. It features a concise place-based encoding to aid in state canonization and is based on the functional module \verb|BAG{X}|, which introduces multisets as a complex data type. Specifically, the commutative and associative \verb|_+_| operator provides an intuitive way to describe a multiset as a weighted sum, for instance, \verb|3 . a + 1 . b|. The sort \verb|Pbag| contains multisets of places.

Each place label (a term of sort \verb|Plab|) is a non-empty list of pairs built of \verb|String| and a \verb|Nat|. Places are uniquely identified by their labels. These pairs represent a symmetric component within a nested hierarchy. Compositional operators annotate places incrementally from right to left: The label suffix represents the root of a hierarchy. 
For example, the 'assembly' place of line 1 in Production Line 2 would be encoded as:
\begin{center}
\verb|p(< "a"; 0 > < "L"; 1 >)|.     
\end{center}

\medskip
We implicitly describe net transitions (\verb|Tran| terms)  through their incidence matrix (a 3-tuple of \verb|Pbag| terms) and associated tags. A tag includes a \verb |String|, a \verb |Nat| (indicating a priority) and a \verb |Float| (interpreted as a firing rate or a probabilistic weight, depending on whether the priority is zero or greater).. 
\[
\texttt{[I,O,H]  |-> << S, P, R >>} 
\]

When using the associative composition operator \verb"_;_" and the subsort relation \verb"Tran < Net", it becomes easy to construct nets in a modular way. For example, we can depict the subnet containing transitions $ld$ and $ln_0$ in Figure \ref{fig:FMS} (top) as the \verb"Net" term in the listing \ref{lst:l1} (the zero-arity operator \verb"nilP" represents an empty multi-set).

\begin{lstlisting}[frame=single,caption=A (sub)net \label{lst:l1},language=maude]
  [2 . p(< "s" ; 0 >), 1 . p(< "w" ; 0 >) + 1 . p(< "w" ; 1 >), nilP] |->  << "ld", 0, 0.5 >> ;
  [1 . p(< "w" ; 0 >), 1 . p(< "a" ; 0 >), 1 . p(< "f" ; 0 > ] |-> << "ln", 0, 0.1 >>
\end{lstlisting}

A \verb|System| term is the empty juxtaposition (\verb|__|) of a \verb|Net| and a \verb|Pbag| (representing the net's marking). 
The conditional rewrite rule \verb|firing| specifies the PT firing rule
\footnote{Notice the use of a matching equation: The free variables T, N',
are matched (:=) against the canonical ground term bound to the variable N.}, as shown in the listing \ref{lst:l2}.

\begin{lstlisting}[frame=single,caption=PT Firing Rule \label{lst:l2},language=maude]
 vars N N' : Net .
 vars T  : Tran .
 var M : Pbag .
 crl [firing] : N M => N fire(T, M) if T ; N' := N  /\ enabled(T, N M) . 
\end{lstlisting}

The predicate \verb|enabled| takes priority into account and relies on \verb|hasConcession|, which determines the 'topological' aspect of the enabling condition:

\begin{lstlisting}[frame=single,caption=PT Firing operators \label{lst:l2-bis},language=maude]
vars I O H M : Pbag . var L : Tlab . 
op hasConcession : Tran Pbag  -> Bool .
eq hasConcession([I,O,H] |-> L, M) = I <= M and-then H > B .
op fire : Tran Pbag -> Pbag .
eq fire([I,O,H] |-> L, M) = M - I + O .
\end{lstlisting}

\noindent A RwSPT is defined by a system module that contains two constant operators, used as aliases: 

\verb|op net : -> Net|

\verb|op m0 :  -> Pbag|.

\medskip
\noindent Two equations define their bindings to concrete terms.
This module includes \verb|System| rewrite rules $R$ incorporating \verb|firing|.
In this paper, we adopt full non-determinism (interleaving semantics): Rewrites take the same priority and have an exponential rate (specified in the rule label but for \verb|firing| rule), so that for the state transition system it holds ($\subseteq$ means subgraph):

$TS($\verb|net m0|, $\{$\verb|firing|$\}) \subseteq TS($\verb|net m0|, $R)$.

\medskip
Transitioning between the \texttt{Maude} encoding of PT systems and the PNML format adopted by many PN tools is straightforward and reversible.

\subsection{Modularity, symmetries, and lumpability} 
We have provided net-algebra and net-rewriting operators \cite{CAPRA-TCS2024} with a twofold intent: to ease the modeler's task and to enable the construction and modification of large-scale models with nested components by implicitly highlighting their symmetry.
A compact \emph{quotient} TS is built using simple manipulation of node labels.
This approach outperforms similar ones, including ours integrated into \texttt{Maude} \cite{Capra:RP22} and based on traditional graph canonization.

In a context where nets have a mutable structure, identifying behavioral equivalences reduces to a graph \emph{morphism}. PT system morphism must maintain the edges and the marking: In our encoding,
a \emph{morphism} between PT systems \verb|(N m)| and \verb|(N' m')| is a bijection $\phi \ :$ \verb|places(N)| $\rightarrow$ \verb|places(N')| such that, considering the homomorphic extension of $\phi$ on multisets, $\phi($\verb|N|$) = \ $\verb|N'| and $\phi($\verb|m|$) = \ $\verb|m'|.
Moreover, $\phi$ must retain the textual annotations of the place labels and the transition tags. If \verb|N'| = \verb|N| we speak of \emph{automorphism}, in which case $\phi$ is a permutation in the set of places.

We refer to a \emph{normal} form that principally involves identifying sets of automorphic (permutable) places:
Two markings \verb|m|, \verb|m'| of a net \verb|N| are said automorphic if there is an automorphism $\phi$ in \verb|N| that maps \verb|m| into \verb|m'|.
We denote this \verb|m| $\cong$ \verb|m'|. The equivalence relation $\cong$ is a congruence, that is, it preserves the transition firings and \emph{rates}.
The next definition helps us simplify the process.
\begin{definition}[Symmetric Labeling]
\label{def:modsym}
A \verb|Net| term is symmetrically labeled if any two maximal sets of places whose labels have the same suffix (possibly empty), which is preceded by pairs with the same tag, are permutable. A \verb|System| term is  symmetrically labeled if its \verb|Net| subterm is.
\end{definition}

\noindent In other words, if a \verb|Net| term \verb|N| meets definition \ref{def:modsym}, then for any two maximal subsets of places matching:

$P := \{$\verb|p(L' < w ; i > L)|$\}, \quad P' := \{$\verb|p(L'' < w ; j > L)|$\}$,

where \verb|L, L', L'' : Plab, w: String, i, j : Nat| 

\noindent there exists an automorphism (permutation) $\phi$ such that $\phi(P) = P'$, $\phi(P') = P$,
which is extended as an identity to the rest of places\footnote{According to the definition of PT morphism, the prefixes \texttt{L'} and \texttt{L''} are consistent in the textual component.}.

If a \verb|System| term adheres to the previous definition, it can be transformed into a 'normal' form by merely swapping indices on the place labels (e.g., i $\leftrightarrow$ j), while still complying with definition \ref{def:modsym}. This normal form is the most minimal according to a lexicographic order within the automorphism class ($\cong$) implicitly defined by \ref{def:modsym}. However, in contrast to general graph canonization, there is no need for any pruning strategy or backtracking. In simple terms, a monotone procedure is used where the sequence of index swaps does not matter (see \cite{CAPRA-TCS2024} for full details). Efficiency is achieved as the normalized form of the subterm of type \verb|Net| is derived through basic ``name abstraction``, where at each hierarchical level the indices in the structured place labels continuously span from $0$ to $k$, $k \in \Nat$.  

The strategy involves providing a concise set of operators that preserve nets' symmetric labelling. This set includes \emph{compositional} operators (influenced by process algebra) and operators for \emph{manipulating} nets, such as adding/removing components.
Rewrite rules require these operators to manipulate \verb|System| terms defined in a modular manner. 
Additionally, rules must adhere to parametricity conditions (here omitted) that limit the use of non-variable terms in them. We denote this kind of rules as symmetric \cite{CAPRA-TCS2024}.

\paragraph{Lumpability} Under these assumptions, we get a \emph{quotient} TS from a \verb|System|
term that
retains reachability and
meets strong bisimulation. 

Let $t,t',u,u'$ be (final) ground terms of sort \verb|System|, 
and let $r$ be a \verb|System| type rule $r : \ s \Longrightarrow s'$. The notation $t \overset{r(\sigma)}{\Longrightarrow} t'$ means that $t$ is rewritten into $t'$ by $r$, that is, there exists a ground substitution $\sigma$ of $r$'s variables such that $\sigma(s) = t$ and $\sigma(s')= t'$.
\begin{property}
\label{prop:transition-corr}
Let
$t$ meet Definition \ref{def:modsym} and $r$ be a symmetric rule.

If $t \overset{r(\sigma)}{\Longrightarrow} t'$ then $\forall u, \phi, t \cong_{\phi} u$: $ u \overset{r(\phi(\sigma))}{\Longrightarrow} u'$, $ t' \cong u'$ ($u', t'$ meet the definition \ref{def:modsym})

\end{property}

The TS quotient produced by a term $\hat{t}$ (pre-normalized) is achieved by applying the overloaded operator \verb|normalize| to the right-hand side of the rewriting rules: 

\verb|op normalize : System -> System .| 

\verb|op normalize : Pbag -> Pbag .|

When a \verb|System| is rewritten using the rule  \verb|firing|, only the marking subterm is needed. This implies applying the overloaded operator \verb|normalize| to the subterm \verb|fire(T, M)|  in Listing 1.2.

According to property \ref{prop:transition-corr}, because the morphism (index exchange) $\phi$ preserves the transition rates and we assume that the rules are parameterized, it is feasible to map the TS quotient of $\hat{t}$ onto an isomorphic "lumped" CTMC: In a Markov process's state space, an equivalence relation is considered "strong lumpability" if the cumulative transition rates between any two states within a class to any other class remain consistent. Despite the possibility of establishing a more stringent condition
matching strong-bisimulation, that is "exact lumpabability" \cite{Bucholz94}, our attention is focused on the aggregated probability.

\paragraph{Example}

To demonstrate the aforementioned concepts, we will outline the compositional RwPT model of a distributed production system with graceful degradation (Section \ref{sec:exe}). Initially, this system is composed of $N$ Production Lines (PL) that share raw materials, with each PL split into $K$ interchangeable branches (listing \ref{lst:l3}).
We start by defining the net transitions. Then we build a Production Line using the \verb"repl&share" operator: The term \verb"PL(K)" represents a Production Line (PL) with \verb"K" symmetric branches, similar to the one shown in Figure 1 (top). The structure of the submodel is expressed by adding a pair with the tag \verb|"L"| to the place labels. For example, \verb|p(< "w" ; 0 > < "L" ; 1 >)| describes the "working" place of robot (line) 1 of the PL. We can also choose to exclude places to share among replicas: In this case, we exclude those representing the "warehouse" (tag "s") and faults (tag "o"). Additionally, we can indicate transitions to share: For instance, "load" and "assembly" are shared.

\begin{lstlisting}[frame=single,,caption=Modular Specification of a Fault Tolerant Production System \label{lst:l3},language=maude]
fmod FTPL is
 pr NET-OP{SPTlab} .
 ops PL PLA nomPL faultyPL NfaultyPL : NzNat -> Net . 
 op faultySys : NzNat -> System . 
 op NPL  : NzNat NzNat -> Net [memo]. 
 op NPLsys : NzNat NzNat NzNat -> System . 
 ops loadLab asLab failLab workLab : -> Tlab [memo] .
 eq loadLab = << "ld",0, 0.5 >> .
 eq asLab = << "as",0, 2.0 >> .
 eq workLab = << "ln",0, 0.1 >> .
 eq failLab = << "ft",0, 0.001 >> .
 var I  : Nat . 
 vars N K M : NzNat .
 eq line =  [1 . p(< "w" ; 0 >),1 . p(< "a" ; 0 >),1 . p(< "f" ; 0 >) ] |-> workLab .
 eq fault = [1 . p(< "o" ; 0 >) , 1 . p(< "f" ; 0 >), nilP ] |-> failLab .
 eq load  = [1 . p(< "s" ; 0 >) , 1 . p(< "w" ; 0 >) , nilP ] |-> loadLab .
 eq ass   = [1 . p(< "a" ; 0 >) , 1 . p(< "s" ; 0 >)  , nilP ]  |-> asLab .
 eq cycle = load ; line ; ass ; fault .
 eq PL(K) = repl&share(cycle, K, "L", p (< "o" ; 0 >) U p(< "s" ; 0 >), asLab U loadLab) .
 eq NPL(N, K) = repl&share(PL(K), N, "PL", p(< "s" ; 0 >), emptyStlab) . 
 eq NPLsys(N, K, M) = setMark(setMark(NPL(N, K), "o" "PL", 1), "s", K * M) .
 ...
endfm
\end{lstlisting}


The term \verb|NPL(N, K)| of type \verb|Net| consists of \verb|N| PLs, each of which contains \verb|K| branches. This net was generated using the \verb|repl&share| operator, which adds the \verb|"PL"| tag to place labels to indicate an additional nesting level. The sharing mechanism ensures each PL gathers \verb|K| raw pieces. The PT net represented by \verb|NPL(2,2)| can be seen in Figure \ref{fig:symPL-degarde2}, top-right. Furthermore, the term  \verb |NPLsys (N, K, M)| of type \verb|System| is a PT system that holds \verb|K*M| tokens in the "warehouse" place, with a single token in each place tagged with \verb|"o"| to trigger fault occurrences within a PL.
We can build an identical model using the "symmetric" version of the process algebra ALT operator.

\medskip
The \verb|System| term generated using the above operators possesses symmetrical labeling (refer to definition \ref{def:modsym}), and its \verb|Net| subterm has already been normalized. 
Consider, e.g., \verb|NPLsys(2, 2, 1)|.
By triggering the conflicting transitions "load", which are initially enabled, the following two markings (essentially, subterms of the \verb|System| terms) can be obtained:

\begin{itemize}
\item[$m_1$]: \verb|p(< "o"; 0 > < "PL"; 0 >) + p(< "o"; 0 > < "PL"; 1 >) +|

\verb|p(< "w"; 0 > < "L"; 0 > < "PL"; 1 >) + p(< "w"; 0 > < "L"; 1 > < "PL"; 1 >)|

\medskip
\item[$m_2$]: \verb|p(< "o"; 0 > < "PL"; 0 >) + p(< "o"; 0 > < "PL"; 1 >) +| 

\verb|p(< "w"; 0 > < "L"; 0 > < "PL"; 0 >) + p(< "w"; 0 > < "L"; 1 > < "PL"; 0 >)|.
\end{itemize}

These are automorphic (one can be converted into the other by interchanging \verb|< "PL"; 1 >| \(\leftrightarrow\) \verb|< "PL"; 0 >|), but the second marking is the smallest in lexicographic order and hence corresponds to the normalized form.

\medskip
The following rewrite rule  (see the listing \ref{lst:l4}) encapsulates the self-adjustment of a PL with  $K=2$ in response to a fault, enabling it to function in a diminished capacity (refer to figure \ref{fig:symPL-degarde2}). This rule deviates slightly from \cite{CAPRA-TCS2024}, as it is locally activated by a breakdown, leading to a significantly larger TS. Skipping technical details, we point out that the rule meets the parameterity and only employs operators that uphold the definition \ref{def:modsym}, such as \verb | join |, \verb | detach |, \verb | setMark |. 
Therefore, it retains the symmetrical PT labeling (definition \ref{def:modsym}).
The label of the rule contains the exponential rewrite rate as meta-information. There is another rule, not discussed here, that eliminates a faulty and degraded PL from the system.

\medskip
\begin{lstlisting}[frame=single,caption=Rewrite rule of a PL (the label contains the rule's exponential rate) \label{lst:l4},language=maude,label=list:rewrulePL]
vars S S' S'' : Pbag . vars I  J : Nat .
var Sys Sys' : System . var L : Lab .
crl [r1-0.005] : N  S => normalize(join(Sys, setMark(setMark(Sys', "w" "fPL", | match(S', "w") |), "a" "fPL", | match(S', "a") |)))  
if  S'' + 1 . p(< "f" ; J > L < "PL" ; I >) := S /\ N' := nomPL(I) /\ dead (N' S) /\ S' := subag(S'', < "PL" ; I >) /\ Sys := detache(N, N') S'' - S' /\ Sys' := faultySys(notIn(N, "fPL"))  .
\end{lstlisting}

\medskip
With the model-checking facilities of \texttt{Maude} (in this case, the \texttt{search} command), it is possible to formally demonstrate that for any given $N$, the quotient transition system has two absorbing states: Every state comprises a deteriorated PL that contains all $2 \cdot M$ materials (unprocessed, except possibly one). This is equivalent to the command below, which yields the same results as its unconditioned counterpart.

\begin{verbatim}
search NPLsys(N,2,M) =>! F:System such that 
net(F:System) == faultyPL /\ B:Pbag := marking(F:System) /\ 
| match(B:Pbag, "w") | + | match(B:Pbag, "a") | == 2 * M  .    
\end{verbatim}


\section{Obtaining the Lumped CTMC generator from an RwSPT}
\label{sec:CTMC}
The CTMC generator entry $Q[i,j]$ is defined as:
$\sum_{r \in R} \lambda_{r} \cdot |S_{i,j}^r|$, where $\lambda_{r} \in \Re^+$ is a given rate, and $S_{i,j}^r = \{ \sigma \ | \ \hat{t}_i \overset{r(\sigma)}{\Longrightarrow} t_j, \ t_j \cong \hat{t}_j \}$ represents the matches of $r$ resulting in equivalent states. 
Therefore, to obtain the CTMC infinitesimal generator, it is necessary to quantify instances that correspond to a specific state transition.
%
%
%
Our solution uses two operators: the first identifies potential matches for each rule based on the subset of independent variables involved, and the second simulates the rewriting process. These two operators can be "mechanically" defined from the syntax of a rule.

To gain a clearer understanding of the concept, let us examine a simplified scenario that encompasses the vast majority of cases and to which any case can be reduced. We suppose that for every rule $r \in R$:
\begin{enumerate}
\item
 $r$ is "injective", that is, if $t \overset{r(\sigma)}{\Longrightarrow} t' \wedge t \overset{r(\sigma')}{\Longrightarrow} t'$ then $\sigma = \sigma'$,
\item  if $r$ is a conditional rule ($r: s \Longrightarrow s' \ if \  cond$) the condition does not contain any rewrite expressions (taking the concrete form $u \Longrightarrow u'$).
\end{enumerate}
Given these assumptions, it is possible to automatically expand a stochastic RwPT specification to produce a quotient TS. The states in this TS encompass all the information required to build the infinitesimal generator of the lumped CTMC, which is isomorphic to the TS.

Listing \ref{lst:l6}, which is related to the running example, describes a general pattern. To avoid overly technical details of \texttt{Maude} syntax, we outline an operator, \verb|rule|, which encodes any rewriting rules except for \verb|firing| (handled separately for efficiency). This operator defines a \emph{partial} mapping where, given a label (defined using a \verb|Tlab|) and a \verb|System| term, it determines the corresponding term-rewriting if feasible: each rewrite rule is tied to an equation. The operator \verb|ruleApp| builds upon \verb|rule|: 
it computes all potential outcomes of rewriting that term using the rule. It does not execute term normalization. As is typical in \texttt{Maude}, the \verb|ruleApp| definition is optimized via tail-recursion. Lastly, \verb|ruleExe|, which extends \verb|ruleApp|, partitions the results of a rule application to a term into "equivalence classes" (sort \verb|Rset|) through normalization: each class is represented by a pair \verb"System <-| Float", 
that is the aggregate rate towards a normalized state. The operator \verb|ruleApp| serves as the bulk form of \verb|ruleExe|.

\begin{lstlisting}[frame=single,caption=rule encoding for the lumped CTMC \label{lst:l6} ,language=maude,label=list:rewrules]
vars N N' N'' : Net . vars S S' S'' : Pbag . vars I Imin J : Nat .
vars Sys Sys' : System . var L : Lab . var Sp : Pset . var TL : Tlab .

op rule : Tlab System -> [System] . *** one equation for rule
ceq rule(<< "r1",0, 0.005 >>, N S) = join(Sys, setMark(setMark(Sys', "w" "fPL", | match(S', "w") |), "a" "fPL", | match(S', "a") |))   
  if  S'' + 1 . p(< "f" ; J > L < "PL" ; I >) := S /\ N' := nomPL(I) /\ dead (N' S) /\ 
    S' := subag(S'', < "PL" ; I >) /\ Sys := detache(N, N') S'' - S' /\ 
    Sys' := faultySys(minNotIn(N, "fPL"))  . 

ceq rule(<< "r2",0, 0.01 >> , N S) = N'' set(S'' - S', p(< "s" ; 0 >), S[p(< "s" ; 0 >)] + | S' |) 
  if S'' + 1 . p(< "f" ; J > L < "fPL" ; I >) :=  S /\ N' := faultyPL(I) /\  dead(N' S)  /\ 
    N'' := detache(N, N') /\ N'' =/= emptyN  /\  S' := subag(S'', < "fPL" ; I >) .

*** "rule application" (without normalization) 
var SS : Set{System} . var TS : [System] . vars R F : Float .
ops ruleApp : Tlab System -> Set{System} .
eq ruleApp(TL, Sys) =  $ruleApp(TL, Sys, emptySS) .
op $ruleApp : Tlab System Set{System} -> Set{System} .
ceq $ruleApp(TL, Sys, SS) = $ruleApp(TL, Sys, SS U TS) if TS := rule(TL, Sys) /\ TS :: System  /\ not(TS in SS) . 
 eq $ruleApp(TL, Sys, SS) = SS [owise] .
*** "aggregate" rates calculation (with normalization) 
op rulexe : Tlab System -> Rset .
eq rulexe(TL, Sys) = $rulexe(rate(TL), ruleApp(TL, Sys), emptyRset) .
op $rulexe : Float Set{System} Rset -> Rset .
eq $rulexe(F, emptySS, RS) = RS .
ceq $rulexe(F, Sys U SS, RS ; Sys' <-| R) = $rulexe(F, SS, RS ; Sys' <-| R + F ) if Sys' := normalize(Sys) .
eq $rulexe(F, Sys U SS, RS) = $rulexe(F, SS, RS ; normalize(Sys) <-| F) 
  [owise] .
op allRew : System -> Rset [memo] . *** bulk application
eq allRew(Sys) = rulexe(labr1, Sys) U rulexe(labr2, Sys) .
\end{lstlisting}

The excerpt in Listing \ref{lst:l5} illustrates the augmented state representation which contains detailed information on the (normalized) state transition.
The state structure defined by the mixfix constructor \verb|StateTranSys| comprises four fields. The initial pair describes the PT system, while the remaining two fields detail the state transitions caused by the \verb|firing| rule and other rewrites, in that order. As explained, we collect state transitions
(rule applications) that share the normalized target for calculating the aggregated rates. 

Now, let us examine the \verb|firing| rule (Listing \ref{lst:l2}): we rephrase it using two related operators, specifically \verb|enabSet|, which determines the set of transitions enabled in a specific marking (or more broadly, the rule's matches), and \verb|fire|, which identifies the resulting markings (the rule's outcomes for all matches), each linked to its respective cumulative rate.
The method applied for the \verb|firing| rule can be generalized to any rule, including those that are not injective.

The function \verb|toStateTran| transforms the traditional state representation into a structured format that emphasizes cumulative transition rates. The actual implementation of the firing rule and other transformation rules is simple, as their effects are immediately apparent in the enhanced state information.


{\small
\begin{lstlisting}[frame=single,caption=TS encoding for the lumped CTMC \label{lst:l5} ,language=maude,label=list:TSextended]
vars B B' M M' : Pbag .  var N : Net . var  TS : TagSet . var  FS   : Fset . var  RS   : Rset .
var R  : Float .
*** description of a system pointing out (aggregate) state-transition rates
op NET:_ M:_ FIRING:_REW:_ : Net Pbag Fset Rset -> StateTranSys [ctor] .
op toStateTran : System -> StateTranSys . 
eq toStateTran(N M) = NET: N  M: M FIRING: fire(enabSet(N M), M) REW: allRew(N M) .
*** caculates the cumulative firing effect of a net (that is, a set of transitions) 
op fire : Net Pbag -> Fset .
*** definition of fire
***
*** implementation of rewrite rules
rl [firing] : NET: N M: B FIRING: (B' <-| R ; FS) REW: RS => toStateTranSPN(N B') .
*** net rewrites
rl [rew] : NET: N M: B FIRING: FS REW: (Sys <-| R ; RS)  => toStateTranSPN(Sys) .
\end{lstlisting}
}

When considering \verb|toStateTran(NPLsys(2,2,2))|, which aligns with the PT net at the top of Figure \ref{fig:symPL-degarde2}, the resulting quotient TS comprises 295 states compared to the 779 states in the standard TS. The quotient graph's state transitions often correspond to multiple matches. For instance, the initial state (the term above) includes two 'load' instances and four 'fault' instances that lead to markings with identical normal forms. Consequently, the combined rates are $2\cdot0.5$ and $4\cdot0.001$. Equivalent rewrites of the net structure are observed when $N > 2$. 

\subsection{Experimental Evidence}
We conclude by showcasing the experimental validation of the method alongside a straightforward demonstration for calculating standard performance metrics. The results of the final-state location command are shown in Table \ref{tab:perf} (above).
This was carried out using Linux WSL on an 11th-gen Intel Core i5 with 40GB RAM. The state spaces align with those of the corresponding lumped CTMC. It is evident that analysis of large models is achievable by leveraging the model's symmetry. Note that the number of absorbing states in the TS quotient remains unchanged with $N$.
Even though a redundant state representation was used to construct the lumped CTMC directly, the efficiency of the \texttt{Maude} rewriting engine allowed us to estimate a time overhead of no more than $80\%$.
\begin{table}[htbp]
\centering\small
\caption{Ordinary vs Quotient TS of \texttt{NPLsys(N,2,2)} \hspace{2pt} ${}^\dag$ \texttt{search} timed out after 10 h} 
\label{tab:perf}
\begin{tabular}{ |p{0.5cm}||p{2.5cm}p{1.5cm}||p{2.5cm}p{1.5cm}| }
\hline
 & \multicolumn{2}{c||}{Ordinary} & \multicolumn{2}{|c|}{Quotient} \\
\verb|N|   & states(final)  & time (sec) &  states(final) & time (sec) \\
\hline
\hline
1   &    60(2) & 0 &      42(2) & 0\\
\hline
2   &    779(4) & 0.1 &     295(2) & 0.1\\
\hline
3  &   6101(6) & 4.8  &  1059(2) & 0.9\\
\hline
4 &    37934(8) & 69 & 2764(2) & 3.6\\
\hline
5 &   204362(10) & 818 & 5970(2) & 10 \\
\hline
6 &   1000187(12) & 13930 & 11367(2) & 27\\
\hline
7 &  - & ${}^\dag$ & 19775(2) & 65 \\
\hline
8 &  - & ${}^\dag$ &  32144(2) & 186 \\
\hline
9 & - & ${}^\dag$ &  49554(2) & 569 \\
\hline
10 & - & ${}^\dag$ & 73215(2) & 2450 \\ 
\hline
\end{tabular}
\end{table}

According to \cite{CAPRA-TCS2024}, the performance of modular RwPT was evaluated against symmetric nets (SN, previously referred to as well-formed nets) \cite{CHIOLA97}, which are colored Petri nets that produce a symbolic reachability graph (SRG) comparable (in its stochastic extension) to a lumped CTMC.
As N and K values rise, the state aggregation level in modular RwPT drastically surpasses that of SN. For example, with N=10, K=3, and M=3, the state aggregation level is around 45 times greater than SN, and with N=10, K=4, and M=3, it is approximately 200 times greater than SN.
You can replicate the experiments following the guidelines at \url{https://github.com/lgcapra/rewpt/tree/main/modSPT/readme}.

\begin{figure}[htbp]
   \begin{center}    
    \includegraphics[clip,trim={2.1cm 2.7cm 2.2cm 3.2cm},width=0.7\columnwidth]{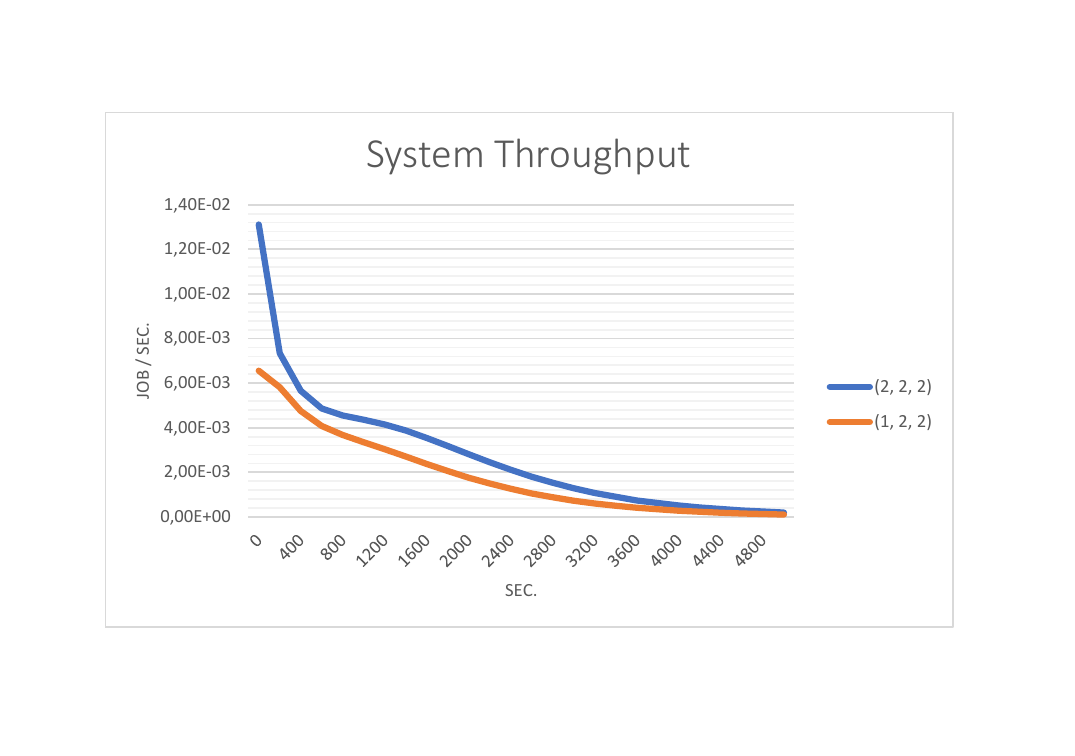}
    \end{center}
    \caption{System Throughput.}
    \label{fig:PlotX}
%
   \begin{center}    
    \includegraphics[clip,trim={2.1cm 2.7cm 2.2cm 3.2cm},width=0.7\columnwidth]{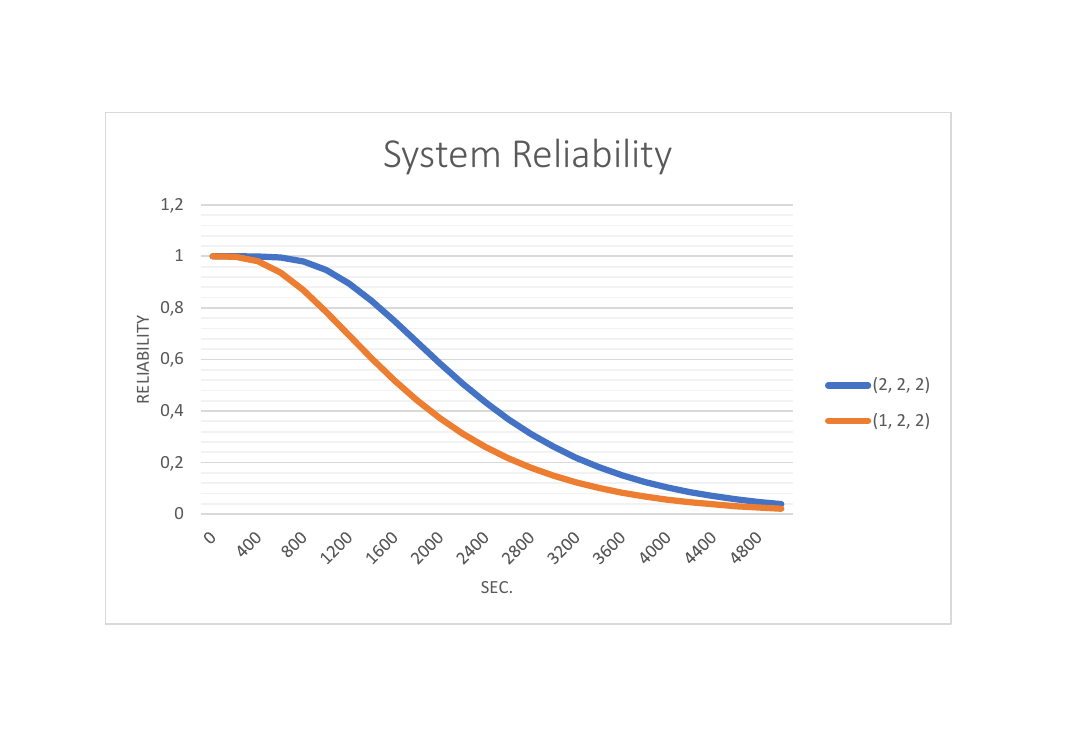}
    \end{center}
    \caption{System Reliability.}
    \label{fig:PlotRel}
%
   \begin{center}    
    \includegraphics[clip,trim={2.1cm 2.7cm 2.2cm 3.2cm},width=0.7\columnwidth]{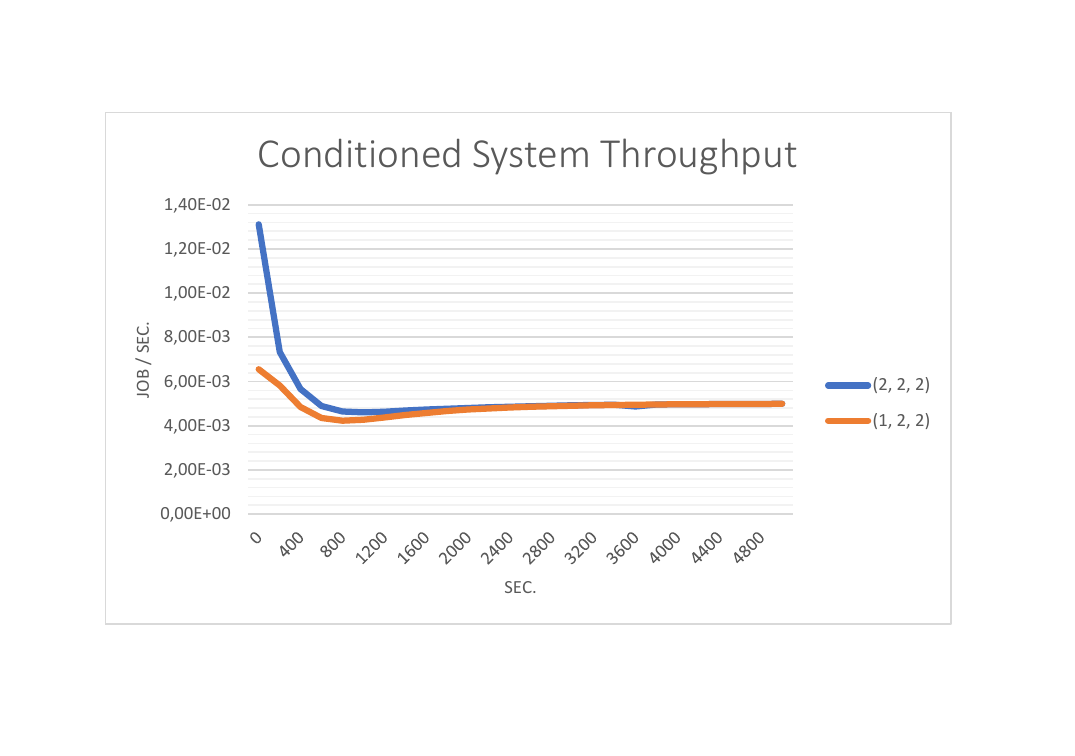}
    \end{center}
    \caption{System Throughput conditioned to its reliability.}
    \label{fig:PlotXdivRel}
\end{figure}

Figure \ref{fig:PlotX} shows the system throughput, while \ref{fig:PlotRel} shows its reliability as a time function. As expected, both metrics decrease with time; additionally, the scenario that involves more replicas demonstrates increased throughput and enhanced reliability.
To evaluate the system's performance, Figure \ref{fig:PlotXdivRel} shows the throughput while the system is operational, which is the ratio between the graphs in Figures \ref{fig:PlotX} and \ref{fig:PlotRel}. It can be seen that the throughput is close to that of a single line, which, given the parameters, is $1/202.5 = 4.98E-03$. The inflection point at around time 800 in both curves represents the system's reconfiguration time. The increased execution time of the job is a result of a system failure.

The overall trend is also noticeable when we look at larger values of \verb|N|. As \verb|N| increases, both reliability and throughput curves show significant improvements. However, we observe an asymptotic trend when \verb|N| is greater than 6. Our interpretation is that beyond a certain point, the benefit of using a higher number of replicas is outweighed by the higher fault rate and the increased configuration overhead.

\section{Conclusion and Future Work}
We have created a Lumped Markov process for modular, rewritable stochastic Petri nets (RwPT), which serves as a robust model for analyzing adaptive distributed systems encoded in \texttt{Maude}. RwPT models, assembled and manipulated through a compact set of (algebraic) operators, display structural symmetries leading to an efficient quotient state transition graph. By providing an example of a gracefully degrading system, we have demonstrated a semi-automatic method for deriving the CTMC infinitesimal generator from the RwPT quotient graph. Future work will, on one hand, delve into exploring orthogonal structured solutions and, on the other, focus on fully implementing the process and integrating it with graphical tools such as \texttt{DrawNET} (\url{https://www.draw-net.com/}). At the same time, we aim to expand the approach: firstly, to derive a lumped Markov process from rewritable GSPN, and secondly, to extract the infinitesimal CTMC generator from any \texttt{Maude} system module.

\bibliographystyle{eptcs}
\bibliography{lc}




\end{document}